\documentclass[12pt]{iopart}
\expandafter\let\csname equation*\endcsname\relax
\expandafter\let\csname endequation*\endcsname\relax
\usepackage{amsfonts}
\usepackage{amsmath}
\usepackage{amssymb}
\usepackage{mathtools}
\usepackage{graphicx}
\usepackage{color}
\usepackage{bbm}
\usepackage{bbold}
\usepackage{hyperref}
\hypersetup{
     colorlinks   = true,
     citecolor    = blue,
     linkcolor    = blue
}
\usepackage{subfigure}
\usepackage{mathptmx}
\usepackage{times}
\usepackage{ulem}

\newcommand{\ignore}[1]{}
\newcommand{\nobibentry}[1]{{\let\nocite\ignore\bibentry{#1}}}

\newcommand{\average}[1]{\left<#1\right>}

\begin{document}
\title{Linear response theory for quantum Gaussian processes}

\author{Mohammad Mehboudi}
\address{ICFO-Institut de Ciencies Fotoniques, The Barcelona Institute of Science and Technology, 08860 Castelldefels (Barcelona), Spain}
\author{Juan M. R. Parrondo}
\address{Departamento de Estructura de la Materia, F\'isica T\'ermica y Electr\'onica and GISC, Universidad Complutense Madrid, 28040 Madrid, Spain}
\author{Antonio Ac\'in$^{1,2}$}
\address{$^{1}$ICFO-Institut de Ciencies Fotoniques, The Barcelona Institute of
Science and Technology, 08860 Castelldefels (Barcelona), Spain}
\address{$^{2}$ICREA-Instituci\'o Catalana de Recerca i Estudis Avan\c cats, 08010, Barcelona, Spain}
\begin{abstract}
    Fluctuation dissipation theorems connect the linear response of a physical system to a perturbation to the steady-state correlation functions. Until now, most of these theorems have been derived for finite-dimensional systems. However, many relevant physical processes are described by systems of infinite dimension in the Gaussian regime. In this work, we find a linear response theory for quantum Gaussian systems subject to time dependent Gaussian channels. In particular, we establish a fluctuation dissipation theorem for the covariance matrix that connects its linear response at any time to the steady state two-time correlations. The theorem covers non-equilibrium scenarios as it does not require the steady state to be at thermal equilibrium. We further show how our results simplify the study of Gaussian systems subject to a time dependent Lindbladian master equation. Finally, we illustrate the usage of our new scheme through some examples. Due to broad generality of the Gaussian formalism, we expect our results to find an application in many physical platforms, such as opto-mechanical systems in the presence of external noise or driven quantum heat devices.
\end{abstract}

\date{\today}
\maketitle
\section{Introduction}
Fluctuation dissipation theorems (FDT) provide very powerful tools to study the linear response of physical systems close to their steady state. 
The aim of such theorems is to establish and quantify a connection between (i) the linear response of the system under study to a (time-dependent) perturbation, and (ii) the steady-state correlation functions.
Different versions of FDT appear, depending on whether the system under study is classical \cite{Cloizeaux,Jensen,Marconi,PhysRev.83.34,Seifert_2010,PhysRevLett.103.090601} or quantum \cite{Kubo_1966,PhysRevX.8.011019,Chetrite2011,konopik2018quantum,Ban2015,Avron_2011,PhysRevA.93.032101}, or whether the steady state is thermal \cite{Kubo_1966}, or a generic non-equilibrium steady state\cite{Seifert_2010,PhysRevX.8.011019,PhysRevLett.103.090601}. 
Response functions have been used to estimate noise \cite{PhysRevD.42.2437,Tapster_1987}, to study topological insulators \cite{PhysRevE.92.052122} or to witness and quantify non-Markovianity of quantum systems \cite{PhysRevLett.121.040601}. 
For a thermal system, or a thermal system subject to a quench, the FDT is connected to the quantum Fisher information \cite{Hauke2016,pappalardi2017multipartite}. On the account of the fact that the quantum Fisher information is a witness of multipartite entanglement \cite{PhysRevA.85.022322,1751-8121-47-42-424006,strobel2014fisher}, one can benefit its connection to the FDT in order to detect multipartite entanglement close to thermal equilibrium. Some recent works report violation of FDTs under certain circumstances \cite{Shimizu_2017,PhysRevB.98.115429}.

To date, the majority of theoretical works in the quantum domain have been focused on finite dimensional systems \textit{and} close to thermal equilibrium. Many physical processes of interest, however, are described by continuous variable systems in the Gaussian regime with infinite-dimensional Hilbert space. These systems also find an application for quantum information technologies and, in fact, have been successfully used for quantum teleportation~\cite{furusawa1998unconditional}, crafting cluster states with enormous number of entangled states~\cite{yokoyama2013ultra} or secure quantum key distribution~\cite{grosshans2003qkd}. It is therefore relevant and timely to establish FDT for quantum Gaussian systems. These theorems should be phrased in terms of the natural tools used to describe Gaussian systems, based on first and second moments rather than density matrices.

In this work we address all these issues and provide a linear response theory for \textit{Gaussian} continuous variable quantum systems.
More specifically, we consider processes described by \textit{Gaussian quantum channels} 
and derive the linear response of the covariance matrix. The formalism can find an application in many different scenarios, since it covers the case of time-dependent fluctuations and non-equilibrium scenarios, as it does not assume the initial state to be thermal. 


The structure of the article is as follows: In section~\ref{sec-definition} we review quantum Gaussian channels. The aim of this section is to provide the minimal necessary tools for this study; for more about Gaussian quantum channels see~\cite{weedbrook2012gaussian,braunstein2005quantum} and the references therein. In section~\ref{sec-Results} we set our framework and present the main results. 
We prove these results in section~\ref{sec-proofs}.
In section~\ref{sec-Lindbladian}, we discuss the application of our theorem for those cases in which the channel is described by a Lindbladian master equation. 
We show how to use our results in section~\ref{sec-examples} through some examples. 
Finally, in section~\ref{sec-Conclusions} we conclude and discuss future directions. 

\section{Definition of the Gaussian scenario}
\label{sec-definition}

We consider $N$-mode Bosonic systems with the quadrature vector $R \equiv (q_1,~\dots q_N;~p_1,~\dots p_N)^T$. Here and throughout the article, $T$ stands for transpose, and the elements $q_i$ and $p_i$ represent the position and momentum of the $i$th mode, respectively. The quadratures respect the Bosonic algebra: $[R_i, R_j] = \Omega_{i,j}$, where $\Omega$ is the symplectic matrix
\begin{align}
	\Omega = i\hbar\left(
	\begin{array}{cc}
	{\mathbb 0}_N &{\mathbbm I}_N\\
	-{\mathbbm I}_N & {\mathbb 0}_N
	\end{array}
	\right).
\end{align}
In our notation, ${\mathbb 0}_N$ is an $N\times N$ matrix of zeros, while ${\mathbbm I}_N$ is the identity matrix of size $N$. In the rest of this work we set $\hbar = 1$ unless otherwise mentioned.
We recall that, by definition, Gaussian systems are those with a Gaussian characteristic function \cite{weedbrook2012gaussian,braunstein2005quantum}. In turn, the characteristic function, denoted by $\chi(\eta)$, reads as:
\begin{align}
\chi(\eta) & \equiv \tr (\rho W_{\eta}) = \tr (\rho {\rm e}^{-\eta^T \Omega R}),
\end{align} 
with $\rho$ being the density matrix, $W_{\eta}\equiv {\rm e}^{-\eta^T~\Omega~R}$ being the Weyl operator, and the phase space vector $\eta$ belongs to ${\mathbbm R}^{2N}$.
Therefore, the characteristic function of a Gaussian system has the following shape:
\begin{align}
\chi(\eta) = {\rm e}^{\frac{1}{2} \eta^T \Omega \sigma \Omega \eta - {\rm d}^T \Omega \eta}.
\end{align} 
Here, we use the displacement vector, denoted by ${\rm d}$, and the covariance matrix, denoted by ${\sigma}$, which are given by:
\begin{align}
	{\rm d} & = {\rm tr}[\rho R],\\
	\sigma_{ij} & = {\rm tr}[\rho~\Sigma_{ij}],
\end{align}
where we define $\Sigma_{ij} \equiv R_i\circ R_j = \frac{1}{2}(\bar{R_i}~\bar{R_j} + \bar{R_j}~\bar{R_i})$, and $\bar{R_i} = R_i - {\rm d}_i$, in order to lighten our notation. The covariance matrix $\sigma$ obeys the uncertainty principle: $\sigma + \Omega/2 \geq 0$ \cite{weedbrook2012gaussian,PhysRevA.49.1567}.

As already advanced, Gaussian systems are fully described by their first and second moments. Thus, Gaussian channels can be completely identified by their action on the displacement vector and the covariance matrix. 
Denoting an arbitrary Gaussian quantum channel by ${\cal M}$, in the most generic case it operates on the quadratures vector and the covariance matrix as follows \cite{weedbrook2012gaussian,braunstein2005quantum,2009arXiv0909.0408H}:
\begin{gather}\label{eq-Gaussian-Channel}
{\cal M}:~{\rm d} \mapsto X {\rm d} + {\rm f},\\
{\cal M}:~\sigma \mapsto X \sigma X^T + Y.
\end{gather}
Here, ${\rm f}\in {\mathbbm R}^{2N}$, while $X$ and $ Y \in {\mathbbm R}^{2N}\times {\mathbbm R}^{2N}$ are real matrices. 
The complete positivity of the map dictates that \cite{weedbrook2012gaussian}:
\begin{align}
 Y + \frac{\Omega}{2} - \frac{1}{2}X \Omega X^T \geq 0.
\end{align} 
Hereafter, without loss of generality, we restrict to zero-mean Gaussian states, and focus on Gaussian channels which map zero-mean states to zero-mean states. This is to say: ${\rm d} = {\rm f} = 0$. On this account, the map ${\cal M}$ could be alternatively characterized by the set $\{X,~Y\}$. In section~\ref{sec-Lindbladian} we review how to bring the particular case of (quadratic) Lindbladian master equations into the standard form of Gaussian channels. 

%

\section{Framework and main results}\label{sec-Results}

We work with a one-parameter family of Gaussian quantum channels ${\cal M}_{\lambda}$, where $\lambda$ is a real parameter, and can represent the strength of an external magnetic field or temperature, to name a few. See Section \ref{sec-examples} for some examples.
Let $\sigma_{\lambda}$ be a fixed point covariance matrix of the Gaussian channel ${\cal M}_{\lambda}$. This implies that:
\begin{align}\label{eq:stationary-criteria}
{\cal M}_{\lambda} \sigma_{\lambda} = \sigma_{\lambda},
\end{align}
or alternatively 
\begin{align}
\sigma_{\lambda} = X_{\lambda}~\sigma_{\lambda}~X_{\lambda}^T + Y_{\lambda}.
\end{align}
Note that we allow both $X_{\lambda}$ and $Y_{\lambda}$ depend on the parameter $\lambda$.
Since we are interested in the linear response, we work in a regime where the parameter $\lambda$ can be considered as a linear contribution to the channel and its fixed point covariance matrix. 
Therefore, we assume that ${\cal M}_{\lambda}={\cal M}_0 + \lambda {\rm M} + {\cal O}(\lambda^2)$, and $\sigma_{\lambda} = \sigma_0 + \lambda\varsigma + {\cal O}(\lambda^2)$, and we safely ignore the second and higher orders. In other words, if we normalize ${\cal M}_0$ and ${\rm M}$ (and doing the same for $\sigma_{0}$ and $\varsigma$) such that they have the same operator norm, then $|\lambda|\ll 1$. Notice that, neither ${\rm M}$ is a Gaussian quantum channel on its own, nor is $\varsigma$ a covariance matrix. We refer to $\varsigma$ as the {\it static} linear response of the covariance matrix, and we have: $\varsigma \equiv
\partial_{\lambda}\sigma_{\lambda}|_{\lambda=0}$. In a Markovian scenario, the time evolution of the covariance matrix is described by the consecutive operations of the map. 
Choosing an initial covariance matrix $\sigma_0$, that is the fixed point of ${\cal M}_0$, at discrete time steps $t=1,2,3,\dots$ we have:
\begin{align}\label{eq:Cov-discrete-evolution}
\sigma(t)={\cal M}_{\lambda(t)}\bullet {\cal M}_{\lambda(t-1)}\dots\bullet{\cal M}_{\lambda(1)} \sigma_0,
\end{align}
where we allow for a {\it time dependent} parameter $\lambda(t)$. The aim is to characterize the linear response of $\sigma(t)$ in terms of steady state correlations, that is elements of steady state covariance matrix.

Our main result expresses the linear response of the covariance matrix as follows:
\begin{align}
\sigma(t) = \sigma_0 + \sum_{s=1}^{t} \lambda(t-s)\Phi(s),
\end{align}
where $\sum$ stands for summation (not to be mistaken with $\Sigma$, the matrix of second order operators), and $\Phi(t)$ is the response function which reads:
\begin{align}\label{eq:response-discrete-C}
\Phi(t) = -\Delta_{t}~\big(X_0^t~\varsigma~{X_0^{T}}^t\big).
\end{align}
Here, $\Delta_t$ stands for time differentiation, i.e., $\Delta_t(f(t)) \equiv f(t+1)-f(t)$ and $X_0^t$ represents $t$ times the application of $X_{\lambda}|_{\lambda=0}$.
If each map is applied for an infinitesimal time $\delta t \to 0$, we have the continuous version of the response function:
\begin{align}
\sigma(t) & = \sigma_0 + \int_0^{t} ds~\lambda(t-s)~\Phi(s),\\
\Phi(t) & = -\partial_t~\big(X_0^t~\varsigma~{X_0^{T}}^t\big). ~\label{eq:response-continious-C}
\end{align} 
Thus, in order to find the response function we simply need to find: (i) the static linear response $\varsigma$, and (ii) the time evolution of $\varsigma$ under the unperturbed channel $X_0$. 
Three further comments/results are in order:

\subsection{Static linear response}
Our result is fully consistent with the static linear response. Consider a scenario in which (i) the perturbation is constant in time, i.e., $\lambda(t)=\lambda $, and (ii) the map has $\sigma_{\lambda}$ as its \textit{unique} fixed point~\footnote{The static linear response holds true only for a scenario in which the steady state is unique. Indeed for the unitary dynamics and thermal states this is not the case, nonetheless our main result---Eq.~\eqref{eq:response-continious-C}---still holds, from which we obtaine the Kubo response.}.
For any arbitrary time $t$ the linear response simplifies to: 
\begin{align}
\sigma(t) -\sigma_0 & = -\lambda\int_0^t ds~\partial_{s}\big(X_0^s~\varsigma~{X_0^{T}}^s\big) + {\cal O}(\lambda^2)\nonumber\\
& = -\lambda~\big(X_0^s~\varsigma~{X_0^{T}}^s\big)_{s=0}^{s=t} + {\cal O}(\lambda^2).
\end{align}
In the limit of $t\to \infty$ the upper value vanishes [see section~\ref{sec-proof-limit}], that is:
\begin{align}\label{eq-limit}
\lim_{s\to\infty}X_0^{s}~\varsigma~X_0^{Ts} = 0.
\end{align}
We immediately revive the static linear response:
\begin{align}
\sigma_{\lambda} - \sigma_0 = \lambda \varsigma + {\cal O}(\lambda^2).
\end{align}
 
\subsection{Kubo's response function}
Consider a system at thermal equilibrium with the density matrix $\rho_0 = {\rm exp}(-\beta H_{0})/Z_{0}$. Here, the quadratic Hamiltonian is given by $H_0 = \frac{1}{2}R^TG~R$, and $\beta$ is the inverse temperature. The system is disturbed by adding a perturbative time-dependent term to its Hamiltonian: $H(t) = H_0 - \lambda(t) h$ with $h = \frac{1}{2}R^Tg~R$. Notice that both $G$ and $g$ are symmetric $2N$ by $2N$ real matrices. The response function of the covariance matrix reads as:
\begin{align}\label{eq-Kubo-final}
\Phi(t) = \frac{i}{\hbar}S_G^{~t}~{\tilde \sigma}~{S_{G}^{T}}^t,
\end{align}
with ${\tilde\sigma} \coloneqq (\Omega g \sigma_0 - \sigma_0 g \Omega)$ and $S_G^{~t} \coloneqq {\rm exp} (-it\Omega G)$. 
Therefore, the linear response of Hamiltonian drivings can be identified only by having $\sigma_0$, $G$ and $g$. 

\subsection{Alternative expression for the linear response}
Practically, Eqs.~\eqref{eq:response-discrete-C} and \eqref{eq:response-continious-C} provide a very useful formalization to find the linear response, for them relying on two elements that are easy to calculate, but they do not seem to follow the usual structure of FDT. However, one can write down an alternative FDT that looks more similar to the traditional one, in particular as presented in \cite{Mehboudi2018fluctuation}. This reads as:
\begin{align}\label{eq:response-SLD}
\Phi(t)=-\partial_t~{\rm Corr} (\Sigma(t),~\Lambda_{0})_0,
\end{align}
where the correlation function is evaluated according to the fixed point of the unperturbed map, hence the index ``0''. Here the elements of the matrix $\Sigma(t)$ are the time evolution of the quadratures under ${\cal M}_0$, such that $\Sigma_{m,n}(t)=R_m(t){\circ}R_n(t)$. In addition, $\Lambda_0$ is the symmetric logarithmic derivative (SLD). The SLD is a Hermitian operator with a vanishing expectation value: $\average{\Lambda_0}_0 = 0$---again the index ``$0$'' indicates that the trace is evaluated over the fixed point of the unperturbed map. In our case it can be written down as a linear combination of second order quadratures:
\begin{align}
\Lambda_0 = \sum_{i,j} C_{ij}\left(R_i \circ R_j - \average{R_i\circ R_j}_0\right),
\end{align}
with the matrix of coefficients $C$ being the solution of the following equation \cite{monras2013phase,nichols2017multiparameter}:
\begin{align}\label{eq-Static-LR-Cov}
\varsigma = 4\sigma_0~C~\sigma_0 + \Omega~C~\Omega.
\end{align} 

\section{Proof of main results}\label{sec-proofs}
Here we present the proof of Eqs.~\eqref{eq:response-discrete-C}, \eqref{eq-limit}, and \eqref{eq-Kubo-final}. The proof of Eq.~\eqref{eq:response-SLD} is presented in the \ref{App:B}.
\subsection{The response function}\label{sec-proof-response}
We start by expanding Eq.~\eqref{eq:Cov-discrete-evolution} and keeping the terms up to the first order in $\lambda$. This yields:
\begin{align}
\sigma(t) &= ({\cal M}_0+\lambda(t){\rm M}) \dots \bullet ({\cal M}_0+\lambda(1){\rm M}) \sigma_0\nonumber\\
& = {\cal M}_0^{t} \sigma_0 + \sum_{s=0}^{t} \lambda(s){\cal M}_0^{t-s} {\rm M} {\cal M}_0^{s-1} \sigma_0\nonumber\\
& = \sigma_0 + \sum_{s=1}^{t}\lambda(s){\cal M}_0^{t-s} {\rm M}\sigma_0\nonumber\\
& = \sigma_0 + \sum_{s=1}^{t}\lambda(s) \Phi(t-s).
\end{align}
In the last line we define the response function $\Phi(t)={\cal M}_0^{t} {\rm M}\sigma_0$.
To proceed further, we need to identify how ${\rm M}$ acts on $\sigma_0$. To this aim, we notice that criterion \eqref{eq:stationary-criteria} implies that:
\begin{gather}
({\cal M}_0+\lambda{\rm M})(\sigma_0 + \lambda \varsigma) = \sigma_0 + \lambda \varsigma + {\cal O}(\lambda^2)\nonumber\\
\Rightarrow {\rm M} \sigma_0 = (1-{\cal M}_0)\varsigma.
\end{gather}
By substituting in the response function, we have:
\begin{equation}\label{eq:phi-sigma1}
\Phi(t) = ({\cal M}_0^t-{\cal M}_0^{t+1})\varsigma\nonumber=-\Delta_t \varsigma(t),
\end{equation}
where we define $\varsigma(t)\equiv {\cal M}_0^t\varsigma$. In section~\ref{sec-definition} we explained how the map ${\cal M}_0$ applies to covariance matrices, however, $\varsigma$ is not a covariance matrix. Thus, we have to identify how the map acts on the static linear response $\varsigma$. 
To this end, we focus on the time evolution of the covariance matrix $\sigma_{\lambda}$:
\begin{align}\label{eq-M0_SLR}
{\cal M}_0^t \sigma_{\lambda} & = X_0^t~\sigma_{\lambda}~{X_0^{T}}^t + \sum_{s=0}^{t-1} X_0^{s}~Y_0~{X_0^{T}}^s 
\nonumber\\
& = X_0^t(\sigma_0 + \lambda \varsigma){X_0^{T}}^t + Y_0(t)+ {\cal O}(\lambda^2)\nonumber\\
& = X_0^t\sigma_0{X_0^{T}}^t + Y_0(t) + \lambda X_0^t\varsigma {X_0^{T}}^t+ {\cal O}(\lambda^2).
\end{align}
Where we define $Y_0(t)\equiv \sum_{s=0}^{t-1} X_0^{s}~Y_0~{X_0^{T}}^s $. By substituting $\sigma_{\lambda} = \sigma_0 + \lambda \varsigma+ {\cal O}(\lambda^2)$ on the left hand side of~\eqref{eq-M0_SLR}, we have:
\begin{align}
{\cal M}_0^t \sigma_0 + \lambda {\cal M}_0^t \varsigma = X_0^t\sigma_0 {X_0^{T}}^t + Y_0(t) + \lambda X_0^t\varsigma {X_0^{T}}^t+ {\cal O}(\lambda^2).
\end{align}
Therefore, we identify the first two terms of the right hand side as ${\cal M}_0^t \sigma_{0}$, and $\varsigma(t) = X_0^t\varsigma {X_0^{T}}^t$. Plugging this into~\eqref{eq:phi-sigma1} completes our proof of Eq.~\eqref{eq:response-discrete-C}.
\subsection{Proof of Equation \eqref{eq-limit}}\label{sec-proof-limit}
The map ${\cal M}_0$ has a unique fixed point $\sigma_0$. This is to say, for any initial covariance matrix $\sigma$ we have:
\begin{equation}
\sigma_0  = \lim_{t\to\infty}{\cal M}_0^t \sigma  = \lim_{t\to\infty} \left(X_0^t \sigma {X_0^{T}}^t + \int_{0}^{t}ds~X_0^s~Y_0~{X_0^{T}}^s\right).  
\end{equation} 
Particularizing to $\sigma_{\lambda} = \sigma_0 + \lambda\varsigma$, yields:
\begin{align}
\sigma_0 & = \lim_{t\to\infty} \left(X_0^t \sigma_{\lambda} {X_0^{T}}^t + \int_{0}^{t}ds~X_0^s~Y_0~{X_0^{T}}^s\right) 
\nonumber\\
& = \lim_{t\to\infty} \left(X_0^t (\sigma_0 + \lambda \varsigma_{\lambda}) {X_0^{T}}^t + \int_{0}^{t}ds~X_0^s~Y_0~{X_0^{T}}^s\right)\nonumber\\
& = \lambda~\lim_{t\to\infty}X_0^t \varsigma {X_0^{T}}^t + \lim_{t\to\infty} \left(X_0^t \sigma_0 {X_0^{T}}^t + \int_{0}^{t}ds~X_0^s~Y_0~{X_0^{T}}^s\right).
\end{align}
Since the equality holds for any value of the parameter, the term proportional to $\lambda$ should vanish, which proves Eq.~\eqref{eq-limit}.
\subsection{FDT for thermal states and Hamiltonian evolutions (Kubo's response function)}
Consider a Gaussian system with $H(t) = H_0 - \lambda(t) h $, where $H_0 = \frac{1}{2}~R^TG~R$, and $h = \frac{1}{2}~R^Tg~R$. 
The thermal state $\rho_{\lambda} = \exp(-\beta H_{\lambda})/{\rm Z_{\lambda}}$---being $Z_{\lambda}$ the partition function---is a fixed point of the unitary dynamics produced by $H_{\lambda}\equiv H_0 - \lambda h$. Recall that, such unitary dynamics corresponds to a symplectic transformation that operates on the covariance matrix (see \ref{a1} for details). In particular we have $X_{0} = {\exp} (-i\Omega G)\eqqcolon S_G$.
By initially preparing the system at the thermal state $\rho_0$ (corresponding to $\lambda = 0$) the linear response reads as:
\begin{align}\label{eq-Kubo-1}
\Phi(t) & = -\partial_t (S_G^{~t}~\varsigma~{S_G^{T}}^t)
= -\partial_t (S_G^{~t}~\partial_{\lambda}{\rm tr}[\Sigma\rho_\lambda]_{\lambda = 0}~{S_G^{T}}^t)\nonumber\\
& = -\partial_t (S_G^{~t}~{\rm tr}\left[\Sigma~(\partial_{\lambda}\rho_{\lambda})_{\lambda = 0}\right]~{S_G^{T}}^t)
= - {\rm tr}\left[\partial_t\Sigma(t)~(\partial_{\lambda}\rho_{\lambda})_{\lambda = 0}\right],
\end{align}
where 
$\Sigma(t) = S_G^{~t}~\Sigma~{S_G^{T}}^t$ is a $2N$ by $2N$ matrix that represents the Heisenberg picture evolution of all of the quadratures. From the Heisenberg equation (or by using Eq.~\eqref{eq-symplectic-shrodinger}) we have:
\begin{align}\label{eq-shrod}
\partial_t \Sigma(t) = -\frac{i}{2\hbar}~[R^TG~R~,~\Sigma(t)].
\end{align}
By plugging \eqref{eq-shrod} into \eqref{eq-Kubo-1}, and using the cyclic property of the trace we have:
\begin{align}\label{eq-Kubo-2}
\Phi(t) & = -\frac{i}{2\hbar}{\rm tr} \left[[\partial_{\lambda}\rho_{\lambda}|_{\lambda = 0}~,~R^TG~R]~\Sigma(t)\right].
\end{align}
To proceed further, we notice that $[H_0 - \lambda h~,~\rho_{\lambda}] = 0$, and hence, by taking the derivative with respect to $\lambda$, and evaluating at $\lambda=0$, we have:
\begin{align}
[\rho_0~,~R^T g~R] = [\partial_{\lambda}\rho_{\lambda}|_{\lambda = 0}~,~R^TG~R].
\end{align}
By inserting the above identity in \eqref{eq-Kubo-2}, and using again the cyclic property of the trace, we have:
\begin{align}\label{eq-Kubo-dm}
\Phi(t) & = \frac{i}{2\hbar}~{\rm tr} \left[[\Sigma(t)~,~R^T g~R]~\rho_0\right],
\end{align}
which can be rewritten in the shape of the standard Kubo-response function:
\begin{align}
\Phi(t) & = \frac{i}{\hbar}~\average{[\Sigma(t)~,~h]}_{0}.
\end{align}
The Kubo response function \eqref{eq-Kubo-dm} can be brought into a more useful shape. To this end, let us look at an individual element of $\Sigma$, say $\Sigma_{lm} = \frac{1}{2}(R_l R_m + R_m R_l)$. The response function of this object has two parts, i.e., $\Phi_{\Sigma_{lm}}(t) = \frac{1}{2}~(\Phi_{R_m R_l}(t) + \Phi_{R_l R_m}(t))$.
We have: 
\begin{align}
\Phi_{R_lR_m}(t) 
& = \frac{i}{\hbar}~\average{[R_l(t)R_m(t)~,~h]}_{\rho_0}\nonumber\\
& = \frac{i}{2~\hbar}\sum_{nn^{\prime}}g_{nn^{\prime}}\average{\left[R_l(t)R_m(t)~,R_nR_{n^{\prime}}\right]}_{0}\nonumber\\
& = \frac{i}{2~\hbar}\sum_{l^{\prime}m^{\prime}nn^{\prime}}g_{nn^{\prime}}(S^{~t}_G)_{ll^{\prime}}~(S^{~t}_G)_{mm^{\prime}}\average{\left[R_{l^{\prime}} R_{m^{\prime}},R_nR_{n^{\prime}}\right]}_{0}\nonumber\\
& = \frac{i}{2~\hbar}\sum_{l^{\prime}m^{\prime}nn^{\prime}}g_{nn^{\prime}}(S^{~t}_G)_{ll^{\prime}}~(S^{~t}_G)_{mm^{\prime}}
\{\Omega_{{m^{\prime}}n^{\prime}}\average{R_{l^{\prime}}R_n}_{0} \nonumber\\
&
+ \Omega_{{m^{\prime}}n}\average{R_{l^{\prime}}R_{n^{\prime}}}_{0}
+ \Omega_{{l^{\prime}}n^{\prime}}\average{R_{n}R_{m^{\prime}}}_{0}
+ \Omega_{{l^{\prime}}n}\average{R_{n^{\prime}}R_{m^{\prime}}}_{0}\}\nonumber\\
& = \frac{i}{\hbar}\{S_G^{~t}\average{RR^T}_0g^T\Omega^T{S_G^{T}}^t + S_G^{~t}~\Omega ~g \average{RR^T}_0{S_G^{T}}^t\}_{lm},
\end{align}
where from first to the second equation we use the definition of $h$, from second to the third we use the fact that for a unitary dynamics $R(t) = S_G^t~R$, from the third to the fourth we expand the commutators and benefit from the canonical commutation relation, and in the last line we reorder everything to show them as product of matrices.
By writing the same expression for $\Phi_{R_mR_l}(t)$, and adding it up to the above result, one obtains:
\begin{align}
	\Phi_{\Sigma_{lm}}(t) & = \frac{i}{\hbar}(S_G^{~t}~\sigma_0~g^T\Omega^T{S_G^{T}}^t~+~S_G^{~t}~\Omega ~g ~\sigma_0~{S_G^{T}}^t)_{lm} = \frac{i}{\hbar}\left(S^{~t}_G{\tilde \sigma}{S_G^{T}}^t\right)_{lm},
\end{align}
with ${\tilde\sigma} \coloneqq (\Omega g \sigma_0 - \sigma_0 g \Omega)$. In a more compact form, for any second order moment we have:
\begin{align}\label{FDR-KUBU}
	\Phi(t) = \frac{i}{\hbar}S^{~t}_G~{\tilde \sigma}~{S_G^{T}}^t.
\end{align}
The equation \eqref{FDR-KUBU} can be considered as the Kubo response function for the covariance matrix of a Gaussian system. Since it is purely defined in terms of the original Hamiltonian ($G$) and the driving force ($g$), it does not require finding $\varsigma$. 
\section{FDT for Lindbladian master equations}\label{sec-Lindbladian}
The stationary state---if it exists---and the elements of the Gaussian channel equivalent to the Lindbladian master equation are found routinely. Let us have a quick reminder about how to formalize this---one could also see for instance \cite{PhysRevA.94.052129}. Consider the following master equation:
\begin{align}
    \frac{d\rho}{dt} = -i[H, \rho] + \sum_{k = 1}^{m}\left(L_k~\rho~L_k^{\dagger}~-~\frac{1}{2}\left\{L_k^{\dagger}~L_k~,~\rho\right\}\right).
\end{align}
Since we are interested in Gaussian dynamics, the Hamiltonian is quadratic in the quadrature operators, and can be written as:
\begin{align}
    H = \frac{1}{2}~R^TG~R,
\end{align}
while the Lindbladian operators can be written as:
\begin{align}
    L_m = c_m^T R,
\end{align}
with $c_m \in \mathbbm{C}^{2N}$ being a vector of size $2N$.
We have ignored some constants in both expressions above, and also a linear dependence of $H$ on the quadratures, since they shall not affect our results significantly.
With these definitions, one can write down the master equation for the covariance matrix and the quadratures vector as follows:
\begin{subequations}
\begin{align}
    \frac{d\average{R}}{dt} & = A \average{R}, \label{eq-Master-Quadrature}\\
    \frac{d\sigma}{dt} & = A \sigma~+~\sigma A^T~+~D, \label{eq-Master-Cov}
\end{align}
\end{subequations}
with the matrix $A = -i\Omega(G-{\rm Im}(C C^{\dagger}))$ and $D = \Omega {\rm Re}(C C^{\dagger}) \Omega$ 
and, the rectangular matrix $C$ is defined as $C = (c_1^T;c_2^T;\dots;c_m^T)^T \in {\mathbbm{C}}^{2N\times m}$ (See the \ref{App:C} or \cite{PhysRevA.85.022103,nicacio2010phase,nicacio2015thermal} for the derivation.) 
From here, the application of the map in an infinitesimal time $\delta t$ can be identified as:
\begin{align}
    \sigma(t) & \mapsto \sigma(t+\delta t) = X(\delta t) \sigma(t) X^T(\delta t) + Y(\delta t),\nonumber\\
    X(\delta t) & = {\rm e}^{\delta t A}, \hspace{1cm} Y(\delta t) = \delta t D 
\end{align}
Finally, the stationary state covariance matrix can be obtained by letting the left hand side of \eqref{eq-Master-Cov} equal to zero. This leads to solving the Lyapunov equation that reads as \cite{HODEL1996205}:
\begin{align}\label{eq-steady-state-Lin}
    A \sigma_{\infty}~+~\sigma_{\infty} A^T~+~D = 0.
\end{align}
The answer to this equation exists and is unique if all of the eigenvalues of $A$ have negative real parts and is given by:
\begin{align}
    \sigma_{\infty} = \int_0^{\infty} dt~{\rm e}^{t A} D ~{\rm e}^{t A^T}.
\end{align}
\section{Examples}\label{sec-examples}

We conclude our study by applying our formalism to two physically relevant examples: a driven harmonic oscillator and a cascaded optimal parametric oscillator.

\subsection{Driven harmonic oscillator}
The thermalization and/or dynamics of quantum open systems is sometimes described in the collision model framework (see for instance~\cite{ziman2005description,PhysRevA.72.022110,lorenzo2017quantum,PhysRevA.96.032107,scarani2002thermalizing}). Following~\cite{eisert2007gaussian} we use the collision model to address the dynamics of a system in presence of thermal noise. The system consists of a single bosonic mode, while the environment consists of an infinite number of bosonic modes. 
The system mode consecutively interacts for some time $\delta t$ with individual environmental modes.
Let $\sigma_E(t)$ denote the state of the environmental mode that interacts with the system at time $t$:
\begin{align}
\sigma_E(t) = c(t) {\mathbbm I}_2.
\end{align} 
Here $(c(t)-2)/2$ represents the mean photon number of the environment mode.
Furthermore, we denote the covariance matrix of the system at time $t$ with $\sigma_s(t)$.
After colliding with the $t/\delta t$th mode of the environment, the covariance matrix of the system maps to:
\begin{align}\label{eq-collision-symplctic}
\sigma_s(t) \mapsto \sigma_s(t + \delta t) = [S_{\eta}(\sigma_s(t) \oplus c(t) {\mathbbm I}_2)~S_{\eta}^T]_{E},
\end{align}
with the index $E$ meaning that we trace out the environmental mode. Moreover, the symplectic matrix $S_{\eta}$ depends on the interaction time $\delta t$ but we have dropped this dependence for lightening our notation. It is given by \cite{eisert2007gaussian}:
\begin{align}
S_{\eta} = 
\left(
\begin{array}{c c}
\sqrt{\eta}~{\mathbbm I}_2 & \sqrt{1-\eta}~{\mathbbm I}_2 \\
-\sqrt{1-\eta}~{\mathbbm I}_2 & \sqrt{\eta}~{\mathbbm I}_2
\end{array}
\right),~~~~ 
\eta \in [0,1].
\end{align}
Here, the parameter $\eta$ quantifies the thermalization rate---which again depends on the interaction length $\delta t$---in particular for $\eta = 0$ the system thermalizes after one application of the map (in this case, the system and the environment exchange their states), whereas for $\eta = 1$ it never thermalizes (in this case, the system and the environment do not interact, and their states remain unchanged). 
By plugging this symplectic transformation into Eq.~\eqref{eq-collision-symplctic}, we can describe the dynamics of the system covariance matrix by means of a quantum Gaussian channel---i.e., with the form of Eq.~\eqref{eq-Gaussian-Channel}. In particular, one can easily check that $X = \sqrt{\eta}~{\mathbbm I}_2$, and $Y = (1-\eta) c(t) {\mathbbm I}_2$. 
We notice that a consecutive application of this map, with fixed $c$ (i.e., $\sigma_E(t) = c_0 \mathbbm{I}_2$, $\forall t$) brings any initial system covariance matrix to the steady state $\sigma_{s}(\infty) = c_0 {\mathbbm I}_2$. In fact, even if the parameter $\eta$ is time dependent, the steady state will remain the same, therefore, we chose it to be constant. Now let us assume that the time dependence of $c(t)$ is a linear correction, i.e, at any time we have $c(t) = c_0 + \lambda(t)$, with $|\lambda(t)|\ll c_0$. Moreover, the initial covariance matrix of the system is $\sigma_s(0) = c_0 \mathbbm{I}_2$.
By using our FDT we aim at identifying the response function. First, notice that $\varsigma = \partial_{\lambda}\sigma_{\lambda}|_{\lambda = 0} = \mathbbm{I}_2$. In addition, since $X = \sqrt{\eta}~\mathbbm{I}_2$, we have $\varsigma(t) = X^{t/\delta t}\varsigma {X^T}^{t/\delta t}~{\mathbbm I}_2 = \eta^{t/\delta t}~{\mathbbm I}_2$. 
Therefore, we have $\Phi(t) = -\partial_t \eta^{t/\delta t}~ {\mathbbm I}_2= -\log(\eta^{1/\delta t}) \eta^{t/\delta t} ~ {\mathbbm I}_2$, which vanishes exponentially in time.

\begin{figure}
	\includegraphics[width=0.9\linewidth]{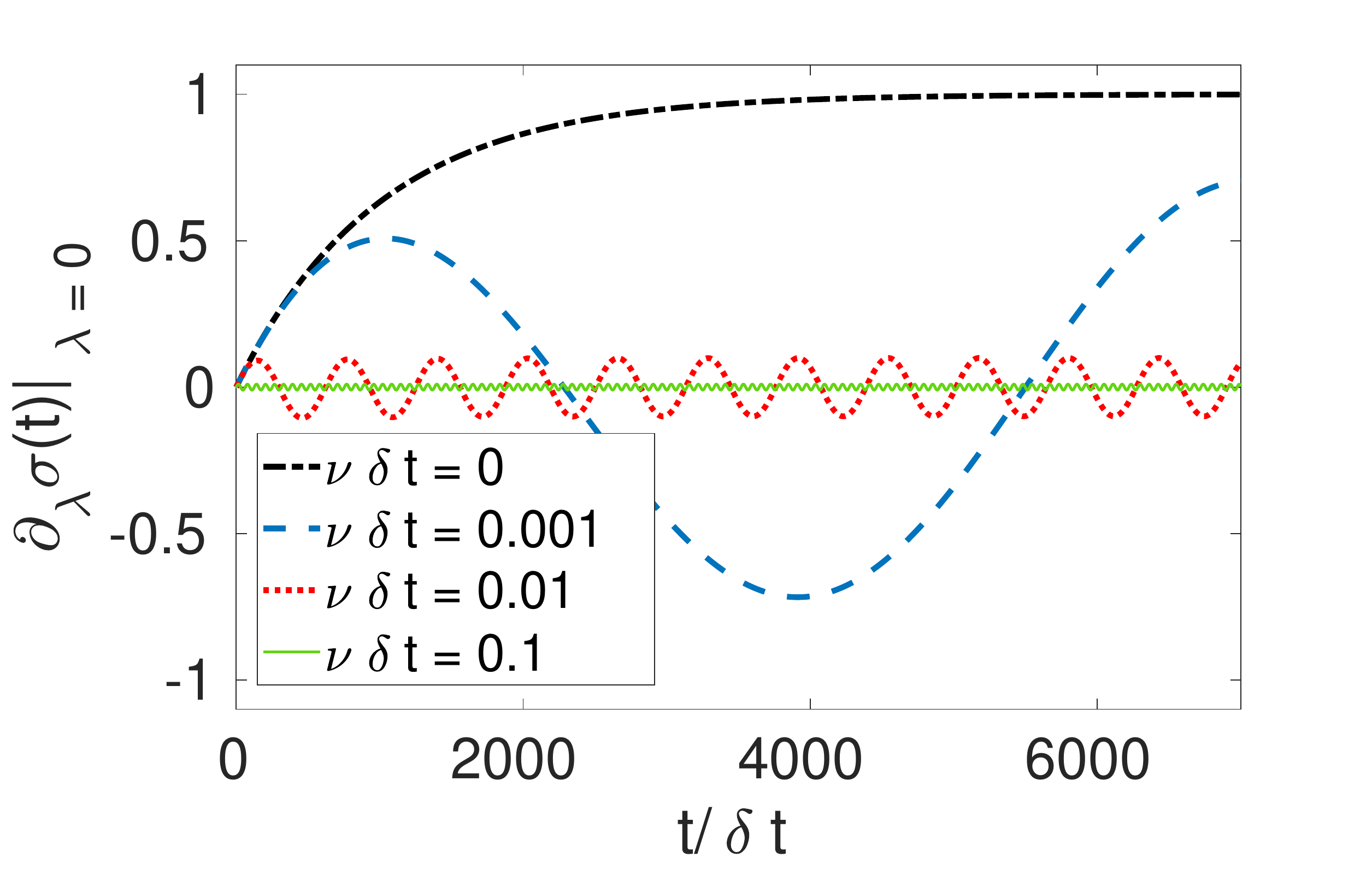}
	\hspace{0.02\linewidth}
	\caption{Linear response ($\partial_{\lambda}\sigma(t)|_{\lambda=0}$) of the covariance matrix for {Example A}. The strength of interaction with the environment is set to $\eta = 0.999$, and $\delta t = 10^{-3}$. The biggest response happens when the perturbation is fixed, i.e., for $\nu = 0$. The response decreases monotonically  by increasing $\nu$, and for big modulation frequency it vanishes.}
	\label{fig-Thermal-Single-O}
\end{figure}

Without any loss of generality let $\lambda(t) =\lambda_0 \cos\nu t$, with $\nu$ being the [potentially tunable] modulation frequency. We shall choose $\nu \delta t \ll 1$, such that the consecutive interaction with the environment is smooth. Specifically, it would be interesting to find the amplitude of the response for different $\nu$. To this aim, 
it is useful to study the linear response of the covariance matrix to the strength of the perturbation:
\begin{align}\label{eq-single-HO-bath}
\partial_{\lambda}\sigma_s(t)|_{\lambda=0}
& ~ = -\int_{0}^{t}\partial_s\eta^{s/\delta t}\cos\nu(t-s)ds~{\mathbbm I}_2\nonumber\\
& ~ = \frac{\tilde{\eta}}{\nu^2 + \tilde{\eta}^2}~\left[\tilde{\eta}\cos\nu t + \nu\sin\nu t - \tilde{\eta}{\rm e}^{-\tilde{\eta} t}\right]{\mathbbm I}_2,
\end{align}    
where we define $\tilde{\eta} = -\log(\eta^{1/\delta t})$ to lighten our notation. In Fig.~\ref{fig-Thermal-Single-O} we depict the linear response, i.e., $\partial_{\lambda}\sigma_s(t)|_{\lambda=0}$ versus time, for different frequencies.
These graphs show how the system responds to a perturbation imposed by manipulating the environment degrees of freedom. Such perturbation can be realized by e.g., changing the temperature or the frequency of the modes in the environment.
The case with $\nu = 0$, specifically illustrates the relaxation of the system to a new thermal state after a quench. 

Indeed, the linear response is initially zero. 
For small $t$, it grows with $\tilde{\eta}t$, regardless of the modulation frequency $\nu$. 
As time increases, the last term in Eq.~\eqref{eq-single-HO-bath} vanishes exponentially. Thus, at long times the system will be oscillating around its initial state unless for the case with $\nu = 0$, i.e., when we have a constant perturbation. The maximum of the response at long times is achieved at $t^*$, the solution of $\tan(\nu t^*) = \nu/\tilde{\eta}$. The value of linear response at such maximum is $\tilde{\eta}/\sqrt{\nu^2+\tilde{\eta}^2}$. Clearly, for any value of $\tilde{\eta}$ the modulation frequency with the biggest linear response corresponds to $\nu = 0$.
Finally, if $\left|\nu/\tilde{\eta}\right|\to \infty $, the response goes to zero, hence the noise is canceled out.
\subsection{Cascaded optical parametric oscillator}
An optical parametric oscillator (OPO) coupled to a vacuum field is described by the Hamiltonian $H = i\epsilon(a^{\dagger 2} - a^2)/4$, with $\epsilon\geq 0$ denoting the effective pump intensity. Here, we use the standard definition of the annihilation and creation operators, that read as $a = (x+i~p)/\sqrt{2}$ and $a^{\dagger} = (x-i~p)/\sqrt{2}$, respectively. The coupling to the vacuum is described by the Lindbladian operator $L = \sqrt{\kappa}~a$, with $\kappa>0$ being the damping cavity rate. 
For this system, it is not difficult to see that the operators $G$ and $C$ read as follow:
\begin{align}
    G = \frac{\epsilon}{2}\left(
    \begin{array}{cc}
        0 & 1\\
        1 & 0
    \end{array}
    \right), \hspace{1cm}
    C = \sqrt{\frac{\kappa}{2}}(1~~i)^T.
\end{align}
From here, one can obtain the matrices $A$ and $D$ that appear in Eq.~\eqref{eq-steady-state-Lin}:
\begin{align}
    A = \left(
    \begin{array}{cc}
        \frac{1}{2}(\epsilon - \kappa) & 0 \\
        0 & -\frac{1}{2}(\epsilon + \kappa)
    \end{array}
    \right), \hspace{1cm}
    D = \frac{\kappa}{2}~{\mathbbm{I}_2}.
\end{align}
Thus, the steady state covariance matrix exists if $\kappa > \epsilon $, and reads as $\sigma = \frac{1}{2}\text{diag}(\kappa/(\kappa - \epsilon)~,~\kappa/(\kappa + \epsilon))$.

The cascaded OPO consists of two interacting optical oscillators with local Lindbladian operators. The Hamiltonian reads as $H = H_1 + H_2 +i (L_1^{\dagger}L_2~-~L_1L_2^{\dagger})/2$, with $H_j = i\epsilon_j(a_j^{\dagger 2} - a_j^2)/4$, and $L_j = \sqrt{\hbar \kappa} a_j$. In turn the dissipation is given by a single operator $L = L_1 + L_2$~
For convenience, in what follows we work in a representation, where the quadrature vector is represented by $R = (x_1~~\dots~~x_N~~p_1~~\dots~~p_N)^T$. 
The matrices $G$ and $C$ read as:
\begin{align}
    G & = \frac{1}{2}\left(
    \begin{array}{cccc}
        0 & 0 & \epsilon_1 & -\kappa \\
        0 & 0 & \kappa & \epsilon_2\\
        \epsilon_1 & \kappa & 0 & 0\\
        -\kappa & \epsilon_2 & 0 & 0
    \end{array}
    \right),
    \nonumber\\
    C & = \sqrt{\frac{\kappa}{2}}(1~~1~~i~~i)^T.
\end{align}
Therefore, one can find the matrices $A$ and $D$ as follow:
\begin{align}
    A & = \left(
    \begin{array}{cc}
        \frac{\epsilon_1 - \kappa}{2} & 0 \\
        -\kappa & \frac{\epsilon_2 - \kappa}{2}
    \end{array}
    \right)\oplus\left(
    \begin{array}{cc}
        -\frac{\epsilon_1 + \kappa}{2} & 0 \\
        -\kappa & -\frac{\epsilon_2 + \kappa}{2}
    \end{array}
    \right),
    \nonumber\\
    D & = \frac{1}{2}\left(
    \begin{array}{cc}
        \kappa & \kappa \\
        \kappa & \kappa
    \end{array}
    \right)\oplus \frac{1}{2} \left(
    \begin{array}{cc}
        \kappa & \kappa \\
        \kappa & \kappa
    \end{array}
    \right).
\end{align}
\begin{figure}
    \includegraphics[width=1\linewidth]{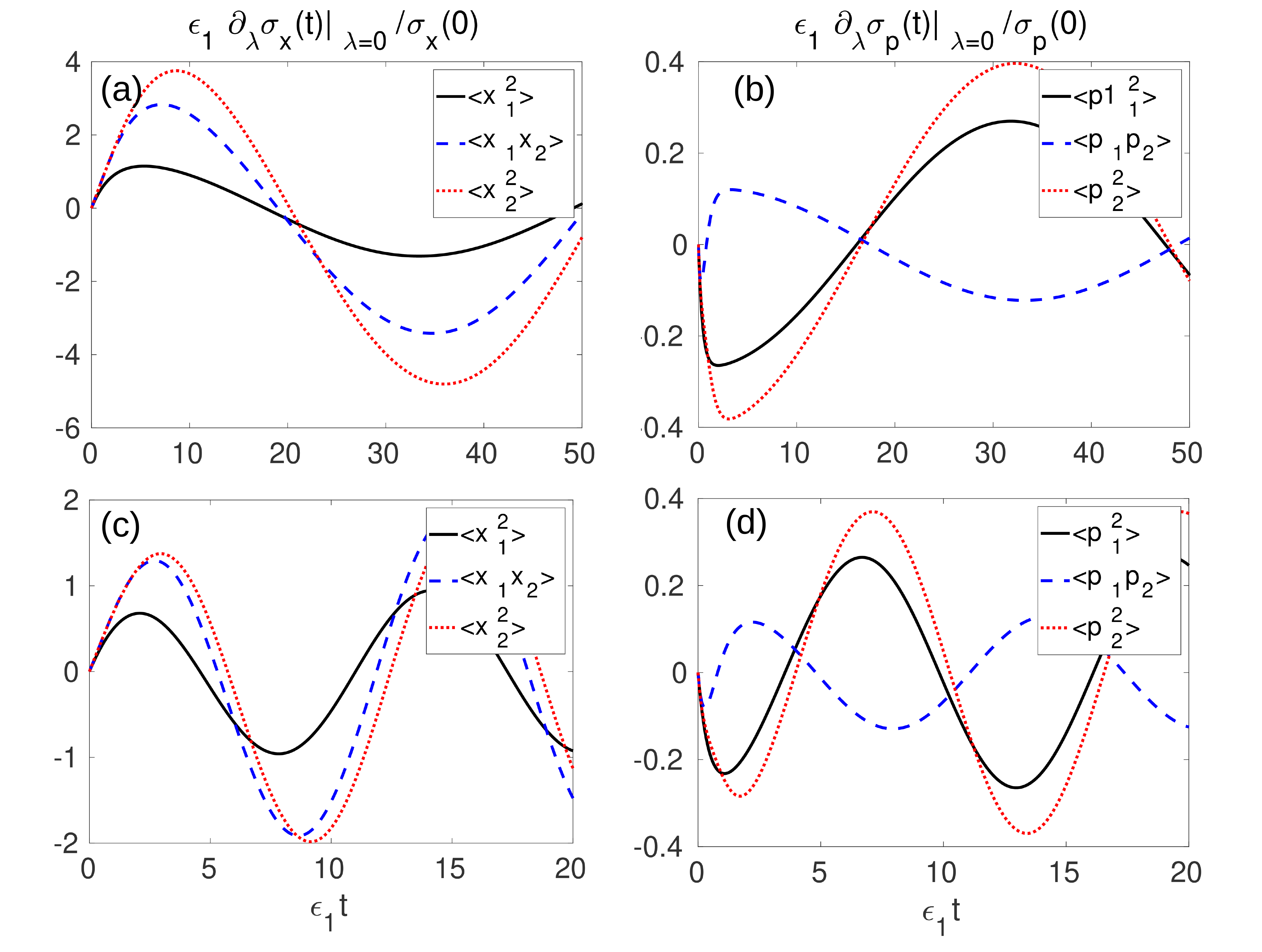}
    \caption{The linear response---normalized by the steady state expectation values---of the position and momentum blocks of the covariance matrix for {Example B}. We consider two different values of modulation frequency $\nu = 0.1$ (top), and $\nu = 0.5$ (bottom). 
    Here we have set the parameters $\epsilon_1 = 1, \epsilon_2 = 1.1$, and the coupling $\kappa$ is derived with the profile $ \kappa = 1.5 + \lambda\cos(\nu t)$.
    Overall, the position quadratures have a bigger response than the momentum ones. In particular, the position of the second oscillator is the most sensitive one. Moreover, it is seen that the bigger frequency $\nu = 0.5$ leads to a smaller response. In fact, for all observables except $\average{p_1p_2}$, the linear response monotonically decreases with increasing the frequency. See Fig.~\ref{fig-Susceptibility-Cascaded-OPO}.
    }
    \label{fig-LR_Cascaded_OPO}
\end{figure}
Again, the criterion for having a steady state is $\kappa>\max\{\epsilon_1~,~\epsilon_2\}$~\cite{PhysRevA.94.052129}.\\ \noindent
Before trying to solve the Lyapunov equation, we note that the matrices $A$ and $G$ are block diagonal, and can be solved in the corresponding blocks. Therefore, the resulting covariance matrix will be a direct sum of two different terms \cite{PhysRevA.94.052129} (one is completely in position subspace, the other one in the momentum subspace, while there are no correlations between the two).
In particular, we need to find the exponential of non-Hermitian matrices $A_x $ and $A_p$, the position and momentum subspaces of the matrix $A$, respectively. To this end, we use the Jordan canonical form of the matrices, that is $A_{\alpha} = V_{\alpha} J_{\alpha} V^{-1}_{\alpha}$. After doing some straightforward algebra, one finds:
\begin{align}
    {\rm e}^{A_x} 
    & = \left(
    \begin{array}{cc}
        {\rm e}^{\frac{\epsilon_1 - \kappa}{2}} & 0 \\
        \frac{-2\kappa}{\epsilon_1 - \epsilon_2}\left( {\rm e}^{\frac{\epsilon_1 - \kappa}{2}} - {\rm e}^{\frac{\epsilon_2 - \kappa}{2}} \right) & {\rm e}^{\frac{\epsilon_2 - \kappa}{2}}
    \end{array}
    \right),
\end{align}
and
\begin{align}
    {\rm e}^{A_p} = & 
    \left(
    \begin{array}{cc}
        {\rm e}^{-\frac{\epsilon_1 + \kappa}{2}} & 0 \\
        \frac{2\kappa}{\epsilon_1 - \epsilon_2}\left( {\rm e}^{-\frac{\epsilon_1 + \kappa}{2}} - {\rm e}^{-\frac{\epsilon_2 + \kappa}{2}} \right) & {\rm e}^{-\frac{\epsilon_2 + \kappa}{2}}
    \end{array}
    \right).
\end{align}
Thus we have the dynamics element for our FDT. Moreover, by putting in the expression of the CM, one finds \cite{PhysRevA.94.052129}:
\begin{align}
\label{eq:covmat-Casopo}
    \sigma = \frac{1}{2}\left(
    \begin{array}{cc}
        \frac{\kappa}{\kappa - \epsilon_1} & \frac{-2\kappa\epsilon_1}{g_-} \\
        \frac{-2\kappa\epsilon_1}{g_-} & \frac{-\kappa h_+}{g_-}
    \end{array}
    \right)\oplus\frac{1}{2}\left(
    \begin{array}{cc}
        \frac{\kappa}{\kappa + \epsilon_1} & \frac{2\kappa\epsilon_1}{g_+} \\
        \frac{2\kappa\epsilon_1}{g_+} & \frac{\kappa h_-}{g_+}
\end{array}
    \right),
\end{align}
where we define $g_{\pm} = (\epsilon_1~+~\epsilon_2~\pm~2\kappa)(\epsilon_1~\pm\kappa)$, and $h_{\pm} = (\epsilon_1^2~+~\epsilon_1~\epsilon_2~\pm~\epsilon_1~\kappa~+~2\kappa^2~\mp\kappa~\epsilon_2)(\epsilon_2~\mp~k)^{-1}$.
This state is always entangled~\cite{PhysRevA.94.052129}.
By making use of the above covariance matrix we can find the other necessary element for our FDT, namely $\varsigma = \partial_{\lambda}\sigma_{\lambda}|_{\lambda = 0}$. 
In turn, the response reads as:
\begin{align}
	\Phi(t) = -\partial_t \left[X^{t}_{0}~\varsigma~{X_{0}^{T}}^t\right].
\end{align}
Notice that, since all the matrices involved in the above equation are block-diagonals, the response will be block-diagonal as well. Therefore, we deal with response function in the position and momentum blocks separately. 
For instance, $\Phi_x(t)$ read as:
\begin{align}
	\Phi_x(t) & = -\partial_t \Big[
	{\rm e}^{A_x t}|_{\lambda=0}(\partial_{\lambda}
	\sigma_x)_{\lambda=0}~
	{\rm e}^{A_x^T t}|_{\lambda=0}
	\Big],
\end{align} 
with $\sigma_x$ being the first matrix in the rhs of Eq.~\eqref{eq:covmat-Casopo}.
Setting $\kappa$ as the driving parameter, that is by choosing $\kappa(t) = \kappa + \lambda\cos\nu t$, the linear response for different values of modulation frequency is depicted in Fig.~\ref{fig-LR_Cascaded_OPO}.
Our first observation is that the position quadratures are more responsive to the perturbation, with $\average{x_2^2}$ having the biggest amplitude of oscillations. From a sensing (estimation) point of view, this means that $\average{x_2^2}$ is the most sensitive quadrature measurement in estimation of $\lambda$ (however, one can design measurements which are superposition of different quadratures and perform even better than $\average{x_2^2}$). We further notice that the response to a perturbation with bigger $\nu$ is smaller---except for the $\average{p_1p_2}$. In fact we can prove that, after long enough time, the amplitude of oscillations of the linear response monotonically decreases with $\nu$. 
\begin{figure}
    \includegraphics[width=1\linewidth]{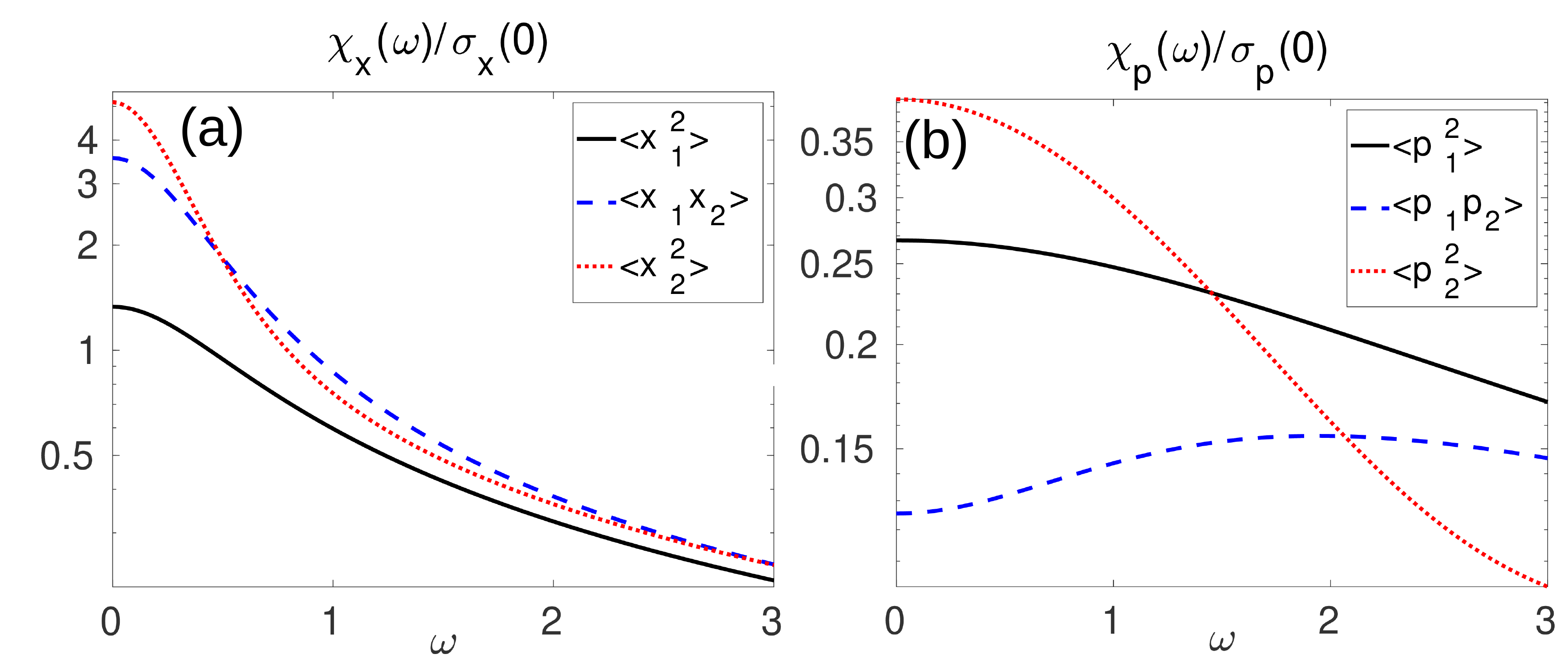}
    \caption{The dynamical susceptibility---normalized by the steady state expectation values---of the position and the momentum parts of the covariance matrix for Example B. 
    For all of the observables---except $\average{p_1p_2}$---the dynamical susceptibility monotonically decreases with the frequency, that explains why the top panel of Fig.~\ref{fig-LR_Cascaded_OPO} has a bigger amplitude of oscillations compared to the bottom panel. Also, the dynamical susceptibility of the position block is significantly larger than that of momentum, which explains why in Fig.~\ref{fig-LR_Cascaded_OPO} position operators have a bigger respond.
    The parameters are the same as in Fig.~\ref{fig-LR_Cascaded_OPO}.}
    \label{fig-Susceptibility-Cascaded-OPO}
\end{figure}
To see this, we recall that the amplitude of oscillations at long times is given by the magnitude of the {\it dynamical susceptibility}. The latter is defined as the Fourier transform of the response function:
\begin{align}
    \chi(\omega) = \int_{-\infty}^{\infty}\Phi(t){\rm e}^{i\omega t} dt 
    = \int_{0}^{\infty}\Phi(t){\rm e}^{i\omega t} dt,
\end{align}
where the integration over negative times is ignored because $\Phi(t<0) = 0$ (due to causality). On the other hand, for any perturbation of the form $\lambda(t) = \lambda\cos\nu t$, the linear response at long times reads as:
\begin{align}
    \partial_{\lambda}\sigma_t|_{\lambda = 0} & \approx \int_{0}^{\infty}\Phi(\tau)\cos\nu(t-\tau)d\tau\nonumber\\
    & = \cos(\nu~t)\int_{0}^{\infty}\Phi(\tau)\cos(\nu\tau)d\tau +\sin(\nu~t)\int_{0}^{\infty}\Phi(\tau)\sin(\nu\tau)d\tau \nonumber\\
    & = \cos(\nu t){\rm Re}\chi(\nu) + \sin(\nu t) {\rm Im}\chi(\nu) = |\chi(\nu)| \cos(\nu t - \alpha),
\end{align}
where we define $\alpha = \cot^{-1}(\frac{{\rm Im}\chi(\omega)}{{\rm Re}\chi(\omega)})$.
Also in the first line, we use the fact that the response function vanishes exponentially at large enough $t$, so that we can replace the upper bound of the integral with infinity. 
Thus, the dynamical susceptibility characterizes the amplitude of oscillations at long times. 
In Fig.~\ref{fig-Susceptibility-Cascaded-OPO} we depict the dynamical susceptibility of position and momentum quadratures. The monotonic decrease in $\chi_{x,p}(\omega)$ with $\omega$ makes it clear why in Fig.~\ref{fig-LR_Cascaded_OPO} we see the amplitude of oscillations decrease with increasing $\omega$. Thus, in an estimation scenario it is more suitable to modulate the perturbation with a small or vanishing frequency. Whereas, if we treat the perturbation as a noise, we shall modulate it with a higher frequency in order to cancel it ot. Comparing the two panels of Fig.~\ref{fig-Susceptibility-Cascaded-OPO} also clarifies why the position quadratures have a considerably larger response than the momentum quadratures.
\section{Conclusions}\label{sec-Conclusions}
We have derived a linear response theory for the covariance matrix of Gaussian systems subjected to time-dependent Gaussian quantum channels.
Our method establishes a connection between the linear response to a time dependent perturbation on the one hand, and on the other hand (i) the {\it static} linear response of the system and (ii) the building blocks of the Gaussian channel itself.
When dealing with thermal states evolving under unitary dynamics, we revive Kubo's linear response theory. We further present an alternative expression for Kubo's response theory, that is more suitable for Gaussian dynamics.
We have then showcased how for any arbitrary (Gaussian) Lindbladian master equation, the two ingredients (i) and (ii) can be identified straightforwardly. Through the examples of thermalization of a harmonic oscillator, and the cascaded parametric oscillator, we have illustrated how to use our formalism. Since Lindbladian master equations appear often in the description of open quantum systems~\cite{weiss2012quantum,breuer2002theory,rivas2012open,wiseman2009quantum}, we expect our results to find an application in many different setups.
In particular, they can be used to improve our understanding of 
opto-mechanical systems~\cite{RevModPhys.86.1391,marquardt2009optomechanics}, 
quantum heat devices~\cite{doi:10.1146/annurev-physchem-040513-103724,PhysRevE.89.042128,PhysRevE.87.042131,skrzypczyk2014work,PhysRevLett.108.070604,PhysRevE.85.061126}, or for the study of the open dynamics of quantum systems in the vicinity of non-thermal steady states.
\section*{acknowledgments}
We thank Anna Sanpera, Andreu Riera-Campeny, and Janek Kolodynski for fruitful discussions. 
Support from the Spanish MINECO (QIBEQI FIS2016-80773-P, ConTrAct FIS2017-83709-
R, and Severo Ochoa SEV-2015-0522), the ERC CoG QITBOX, the AXA Chair in Quantum Information Science, Fundacio Privada Cellex, and the Generalitat de Catalunya (CERCA Program and SGR1381) is acknowledged.
-------------------------
\section*{References}
\bibliographystyle{iopart-num}
\bibliography{Refs.bib}

\providecommand{\newblock}{}
\begin{thebibliography}{10}
\expandafter\ifx\csname url\endcsname\relax
  \def\url#1{{\tt #1}}\fi
\expandafter\ifx\csname urlprefix\endcsname\relax\def\urlprefix{URL }\fi
\providecommand{\eprint}[2][]{\url{#2}}

\bibitem{Cloizeaux}
des Cloizeaux D 1968 {\em Linear response, generalized susceptibility and
  dispersion theory\/} (International Atomic Energy Agency) pp 325--354
  \urlprefix\url{http://www.iaea.org/inis/collection/NCLCollectionStore/_Public/46/031/46031648.pdf#page=334}

\bibitem{Jensen}
Jensen J and Mackintosh A~R 1991 {\em Rare Earth Magnetism Structures and
  Excitations\/} (Clarendon Press)

\bibitem{Marconi}
Marconi U~M~B, Puglisi A, Rondoni L and Vulpiani A 2008 {\em Physics Reports\/}
  {\bf 461} 111 -- 195 ISSN 0370-1573
  \urlprefix\url{http://www.sciencedirect.com/science/article/pii/S0370157308000768}

\bibitem{PhysRev.83.34}
Callen H~B and Welton T~A 1951 {\em Phys. Rev.\/} {\bf 83}(1) 34--40
  \urlprefix\url{https://link.aps.org/doi/10.1103/PhysRev.83.34}

\bibitem{Seifert_2010}
Seifert U and Speck T 2010 {\em {EPL} (Europhysics Letters)\/} {\bf 89} 10007
  \urlprefix\url{https://doi.org/10.1209%2F0295-5075%2F89%2F10007}

\bibitem{PhysRevLett.103.090601}
Prost J, Joanny J~F and Parrondo J~M~R 2009 {\em Phys. Rev. Lett.\/} {\bf
  103}(9) 090601
  \urlprefix\url{https://link.aps.org/doi/10.1103/PhysRevLett.103.090601}

\bibitem{Kubo_1966}
Kubo R 1966 {\em Reports on Progress in Physics\/} {\bf 29} 255--284
  \urlprefix\url{https://doi.org/10.1088%2F0034-4885%2F29%2F1%2F306}

\bibitem{PhysRevX.8.011019}
\AA{}berg J 2018 {\em Phys. Rev. X\/} {\bf 8}(1) 011019
  \urlprefix\url{https://link.aps.org/doi/10.1103/PhysRevX.8.011019}

\bibitem{Chetrite2011}
Chetrite R and Gupta S 2011 {\em Journal of Statistical Physics\/} {\bf 143}
  543 ISSN 1572-9613 \urlprefix\url{https://doi.org/10.1007/s10955-011-0184-0}

\bibitem{konopik2018quantum}
Konopik M and Lutz E 2018 {\em arXiv preprint arXiv:1811.12277\/}
  \urlprefix\url{https://arxiv.org/abs/1811.12277}

\bibitem{Ban2015}
Ban M 2015 {\em Quantum Studies: Mathematics and Foundations\/} {\bf 2} 51--62
  ISSN 2196-5617 \urlprefix\url{https://doi.org/10.1007/s40509-015-0034-x}

\bibitem{Avron_2011}
Avron J~E, Fraas M, Graf G~M and Kenneth O 2011 {\em New Journal of Physics\/}
  {\bf 13} 053042
  \urlprefix\url{https://doi.org/10.1088%2F1367-2630%2F13%2F5%2F053042}

\bibitem{PhysRevA.93.032101}
Campos~Venuti L and Zanardi P 2016 {\em Phys. Rev. A\/} {\bf 93}(3) 032101
  \urlprefix\url{https://link.aps.org/doi/10.1103/PhysRevA.93.032101}

\bibitem{PhysRevD.42.2437}
Saulson P~R 1990 {\em Phys. Rev. D\/} {\bf 42}(8) 2437--2445
  \urlprefix\url{https://link.aps.org/doi/10.1103/PhysRevD.42.2437}

\bibitem{Tapster_1987}
Tapster P~R, Rarity J~G and Satchell J~S 1987 {\em Europhysics Letters
  ({EPL})\/} {\bf 4} 293--299
  \urlprefix\url{https://doi.org/10.1209%2F0295-5075%2F4%2F3%2F007}

\bibitem{PhysRevE.92.052122}
Shen H~Z, Qin M, Shao X~Q and Yi X~X 2015 {\em Phys. Rev. E\/} {\bf 92}(5)
  052122 \urlprefix\url{https://link.aps.org/doi/10.1103/PhysRevE.92.052122}

\bibitem{PhysRevLett.121.040601}
Strasberg P and Esposito M 2018 {\em Phys. Rev. Lett.\/} {\bf 121}(4) 040601
  \urlprefix\url{https://link.aps.org/doi/10.1103/PhysRevLett.121.040601}

\bibitem{Hauke2016}
Hauke P, Heyl M, Tagliacozzo L and Zoller P 2016 {\em Nat Phys\/} {\bf 12}
  778--782 \urlprefix\url{http://dx.doi.org/10.1038/nphys3700}

\bibitem{pappalardi2017multipartite}
Pappalardi S, Russomanno A, Silva A and Fazio R 2017 {\em Journal of
  Statistical Mechanics: Theory and Experiment\/} {\bf 2017} 053104
  \urlprefix\url{http://iopscience.iop.org/article/10.1088/1742-5468/aa6809/meta}

\bibitem{PhysRevA.85.022322}
T\'oth G 2012 {\em Phys. Rev. A\/} {\bf 85}(2) 022322
  \urlprefix\url{https://link.aps.org/doi/10.1103/PhysRevA.85.022322}

\bibitem{1751-8121-47-42-424006}
T\'oth G and Apellaniz I 2014 {\em Journal of Physics A: Mathematical and
  Theoretical\/} {\bf 47} 424006
  \urlprefix\url{http://stacks.iop.org/1751-8121/47/i=42/a=424006}

\bibitem{strobel2014fisher}
Strobel H, Muessel W, Linnemann D, Zibold T, Hume D~B, Pezz{\`e} L, Smerzi A
  and Oberthaler M~K 2014 {\em Science\/} {\bf 345} 424--427
  \urlprefix\url{https://doi.org/10.1126/science.1250147}

\bibitem{Shimizu_2017}
Shimizu A and Fujikura K 2017 {\em Journal of Statistical Mechanics: Theory and
  Experiment\/} {\bf 2017} 024004
  \urlprefix\url{https://doi.org/10.1088%2F1742-5468%2Faa5a67}

\bibitem{PhysRevB.98.115429}
Kubo K, Asano K and Shimizu A 2018 {\em Phys. Rev. B\/} {\bf 98}(11) 115429
  \urlprefix\url{https://link.aps.org/doi/10.1103/PhysRevB.98.115429}

\bibitem{furusawa1998unconditional}
Furusawa A, S{\o}rensen J~L, Braunstein S~L, Fuchs C~A, Kimble H~J and Polzik
  E~S 1998 {\em Science\/} {\bf 282} 706--709 \urlprefix\url{DOI:
  10.1126/science.282.5389.706}

\bibitem{yokoyama2013ultra}
Yokoyama S, Ukai R, Armstrong S~C, Sornphiphatphong C, Kaji T, Suzuki S,
  Yoshikawa J~i, Yonezawa H, Menicucci N~C and Furusawa A 2013 {\em Nature
  Photonics\/} {\bf 7} 982
  \urlprefix\url{https://doi.org/10.1038/nphoton.2013.287}

\bibitem{grosshans2003qkd}
Grosshans F, Van~Assche G, Wenger J, Brouri R, Cerf N~J and Grangier P 2003
  {\em Nature\/} {\bf 421}(6920) 238
  \urlprefix\url{https://www.nature.com/articles/nature01289}

\bibitem{weedbrook2012gaussian}
Weedbrook C, Pirandola S, Garc{\'\i}a-Patr{\'o}n R, Cerf N~J, Ralph T~C,
  Shapiro J~H and Lloyd S 2012 {\em Reviews of Modern Physics\/} {\bf 84} 621
  \urlprefix\url{https://link.aps.org/doi/10.1103/RevModPhys.84.621}

\bibitem{braunstein2005quantum}
Braunstein S~L and Van~Loock P 2005 {\em Reviews of Modern Physics\/} {\bf 77}
  513 \urlprefix\url{https://link.aps.org/doi/10.1103/RevModPhys.77.513}

\bibitem{PhysRevA.49.1567}
Simon R, Mukunda N and Dutta B 1994 {\em Phys. Rev. A\/} {\bf 49}(3) 1567--1583
  \urlprefix\url{https://link.aps.org/doi/10.1103/PhysRevA.49.1567}

\bibitem{2009arXiv0909.0408H}
Heinossari T, Holevo A~S and Wolf M 2010 {\em Quantum Information and
  Computation\/} {\bf 10} 619--635

\bibitem{Mehboudi2018fluctuation}
Mehboudi M, Sanpera A and Parrondo J~M~R 2018 {\em {Quantum}\/} {\bf 2} 66 ISSN
  2521-327X \urlprefix\url{https://doi.org/10.22331/q-2018-05-24-66}

\bibitem{monras2013phase}
Monras A 2013 {\em arXiv:1303.3682\/}
  \urlprefix\url{https://arxiv.org/abs/1303.3682}

\bibitem{nichols2017multiparameter}
Nichols R, Liuzzo-Scorpo P, Knott P~A and Adesso G 2018 {\em Phys. Rev. A\/}
  {\bf 98}(1) 012114
  \urlprefix\url{https://link.aps.org/doi/10.1103/PhysRevA.98.012114}

\bibitem{PhysRevA.94.052129}
Nicacio F, Paternostro M and Ferraro A 2016 {\em Phys. Rev. A\/} {\bf 94}(5)
  052129 \urlprefix\url{https://link.aps.org/doi/10.1103/PhysRevA.94.052129}

\bibitem{PhysRevA.85.022103}
Koga K and Yamamoto N 2012 {\em Phys. Rev. A\/} {\bf 85}(2) 022103
  \urlprefix\url{https://link.aps.org/doi/10.1103/PhysRevA.85.022103}

\bibitem{nicacio2010phase}
Nicacio F, Maia R~N, Toscano F and Vallejos R~O 2010 {\em Physics Letters A\/}
  {\bf 374} 4385--4392
  \urlprefix\url{https://doi.org/10.1016/j.physleta.2010.08.076}

\bibitem{nicacio2015thermal}
Nicacio F, Ferraro A, Imparato A, Paternostro M and Semi\~ao F~L 2015 {\em
  Phys. Rev. E\/} {\bf 91}(4) 042116
  \urlprefix\url{https://link.aps.org/doi/10.1103/PhysRevE.91.042116}

\bibitem{HODEL1996205}
Hodel A, Tenison B and Poolla K~R 1996 {\em Linear Algebra and its
  Applications\/} {\bf 236} 205 -- 230 ISSN 0024-3795
  \urlprefix\url{http://www.sciencedirect.com/science/article/pii/0024379594001553}

\bibitem{ziman2005description}
Ziman M, {\v{S}}telmachovi{\v{c}} P and Bu{\v{z}}ek V 2005 {\em Open systems \&
  information dynamics\/} {\bf 12} 81--91
  \urlprefix\url{https://doi.org/10.1007/s11080-005-0488-0}

\bibitem{PhysRevA.72.022110}
Ziman M and Bu\ifmmode~\check{z}\else \v{z}\fi{}ek V 2005 {\em Phys. Rev. A\/}
  {\bf 72}(2) 022110
  \urlprefix\url{https://link.aps.org/doi/10.1103/PhysRevA.72.022110}

\bibitem{lorenzo2017quantum}
Lorenzo S, Ciccarello F, Palma G~M and Vacchini B 2017 {\em Open Systems \&
  Information Dynamics\/} {\bf 24} 1740011
  \urlprefix\url{https://doi.org/10.1142/S123016121740011X}

\bibitem{PhysRevA.96.032107}
Lorenzo S, Ciccarello F and Palma G~M 2017 {\em Phys. Rev. A\/} {\bf 96}(3)
  032107 \urlprefix\url{https://link.aps.org/doi/10.1103/PhysRevA.96.032107}

\bibitem{scarani2002thermalizing}
Scarani V, Ziman M, \ifmmode \check{S}\else
  \v{S}\fi{}telmachovi\ifmmode~\check{c}\else \v{c}\fi{} P, Gisin N and
  Bu\ifmmode~\check{z}\else \v{z}\fi{}ek V 2002 {\em Phys. Rev. Lett.\/} {\bf
  88}(9) 097905
  \urlprefix\url{https://link.aps.org/doi/10.1103/PhysRevLett.88.097905}

\bibitem{eisert2007gaussian}
{Eisert} J and {Wolf} M~M {2007} {\em World Scientific\/}  {23--42}
  \urlprefix\url{https://doi.org/10.1142/9781860948169_0002}

\bibitem{weiss2012quantum}
Weiss U 2012 {\em Quantum dissipative systems\/} vol~13 (World scientific)

\bibitem{breuer2002theory}
Breuer H~P and Petruccione F 2002 {\em The theory of open quantum systems\/}
  (Oxford University Press)

\bibitem{rivas2012open}
Rivas A and Huelga S~F 2012 {\em Open quantum systems\/} (Springer)

\bibitem{wiseman2009quantum}
Wiseman H~M and Milburn G~J 2009 {\em Quantum measurement and control\/}
  (Cambridge university press)

\bibitem{RevModPhys.86.1391}
Aspelmeyer M, Kippenberg T~J and Marquardt F 2014 {\em Rev. Mod. Phys.\/} {\bf
  86}(4) 1391--1452
  \urlprefix\url{https://link.aps.org/doi/10.1103/RevModPhys.86.1391}

\bibitem{marquardt2009optomechanics}
Marquardt F and Girvin S~M 2009 {\em arXiv preprint arXiv:0905.0566\/}
  \urlprefix\url{https://arxiv.org/abs/0905.0566}

\bibitem{doi:10.1146/annurev-physchem-040513-103724}
Kosloff R and Levy A 2014 {\em Annual Review of Physical Chemistry\/} {\bf 65}
  365--393
  \urlprefix\url{https://doi.org/10.1146/annurev-physchem-040513-103724}

\bibitem{PhysRevE.89.042128}
Correa L~A 2014 {\em Phys. Rev. E\/} {\bf 89}(4) 042128
  \urlprefix\url{https://link.aps.org/doi/10.1103/PhysRevE.89.042128}

\bibitem{PhysRevE.87.042131}
Correa L~A, Palao J~P, Adesso G and Alonso D 2013 {\em Phys. Rev. E\/} {\bf
  87}(4) 042131
  \urlprefix\url{https://link.aps.org/doi/10.1103/PhysRevE.87.042131}

\bibitem{skrzypczyk2014work}
Skrzypczyk P, Short A~J and Popescu S 2014 {\em Nature communications\/} {\bf
  5} 4185 \urlprefix\url{https://doi.org/10.1038/ncomms5185}

\bibitem{PhysRevLett.108.070604}
Levy A and Kosloff R 2012 {\em Phys. Rev. Lett.\/} {\bf 108}(7) 070604
  \urlprefix\url{https://link.aps.org/doi/10.1103/PhysRevLett.108.070604}

\bibitem{PhysRevE.85.061126}
Levy A, Alicki R and Kosloff R 2012 {\em Phys. Rev. E\/} {\bf 85}(6) 061126
  \urlprefix\url{https://link.aps.org/doi/10.1103/PhysRevE.85.061126}

\end{thebibliography}
\appendix
\section{Symplectic representation of a Gaussian unitary transformation}\label{a1}
Suppose that the density matrix of a Gaussian systems evolves under a unitary transformation $U(t) = {\rm exp}(-itH)$, with the quadratic Hamiltonian $H = \frac{1}{2}R^{T}G~R$. This is to say: $\rho(t) = U(t)\rho(0)U^{-1}(t)$. Under this unitary, in the Heisenberg picture, the quadratures evolve as:
\begin{align}\label{eq-symplectic-shrodinger}
R(t) = {\rm e}^{-it\Omega G}R\coloneqq S_G^{~t}R.
\end{align}
\textit{Proof---}By writing the Heisenberg picture evolution of the $j$th element of the quadrature vector, and using the Baker-Campbell-Hausdorff formula we have:
\begin{align}
R_j(t) & = {\rm e}^{\frac{it}{2} R^TG R} R_j {\rm e}^{\frac{-it}{2}R^TG R}\nonumber\\
& = R_j + \frac{it}{2}[R^TG R,R_j] + \frac{1}{2!}(\frac{it}{2})^2 [R^TG R,[R^TG R,R_j]] \nonumber\\ 
&~+ \frac{1}{3!}(\frac{it}{2})^3 [R^TG R,[R^TG R,[R^TG R,R_j]]] + \dots~\nonumber\\
& = R_j + it(\Omega G R)_j + \left(\frac{(it \Omega G)^2}{2!}R\right)_j + \left(\frac{(it \Omega G)^3}{3!}R\right)_j + \dots\nonumber\\
& = ({\rm e}^{-it\Omega}R)_j \coloneqq (S_G^{~t} R)_j,
\end{align} 
where we use the fact that $[R^TG R,R_j] = (\Omega G R)_j$.
\section{Proof of the alternative shape of the response function}\label{App:B}
On the one hand, by expanding $\varsigma(t)$ with the help of Eq.~\eqref{eq-Static-LR-Cov} we have:
\begin{align}\label{eq-On-the-one-hand}
\varsigma(t) & = X_0^t (4\sigma_0 C \sigma_0 + \Omega C \Omega) {X_0^{T}}^t = 4\sigma^{t0} C (\sigma^{t0})^T - \Omega^{t0} C (\Omega^{t0})^{T},
\end{align}
where we have defined the {\it non-symmetric} two-time correlation matrix $\sigma^{t0}$ with the elements $\sigma^{t0}_{m,r}\equiv \average{R_m(t)\circ R_r}_0$, and $\Omega^{t0}$ with $\Omega^{t0}_{m,r}\equiv \average{\left[R_m(t)~,~R_r\right]}_0$. In turn, $R_m(t)\equiv {\cal M}_0^t R_m = X^{t} R_m$ represents the time evolution of the quadratures vector, under the unperturbed map.
%
On the other hand, with the help of Wick's theorem, one can expand the fourth order moments that appear in the right hand side of Eq.~\eqref{eq:response-SLD}, in terms of second order moments. Specifically---by breaking the elements of $\Sigma(t)$ into two parts as in $\Sigma(t)_{m,n} = \frac{1}{2}~(R_m(t)R_n(t) + R_n(t)R_m(t))$---we have:
\begin{align}
{\rm Corr}(\Lambda_0,R_m(t)R_n(t)) & = \sum_{p,q} C_{p,q} {\rm Corr}\big((R_p\circ R_q - \average{R_p\circ R_q}),R_m(t)R_n(t)\big)\nonumber\\
& = \frac{1}{2}\sum_{p,q} C_{p,q}\big(\average{(R_p\circ R_q - \average{R_p\circ R_q})R_m(t)R_n(t)}\nonumber\\
&~~~~~~~~~~~~~~~~~~+\average{R_m(t)R_n(t)(R_p\circ R_q - \average{R_p\circ R_q})}\big)\nonumber\\
& = \frac{1}{2}\sum_{p,q} C_{p,q}\big(
\average{(R_p\circ R_q) R_m(t)R_n(t)}-\average{R_p\circ R_q}\average{R_m(t) R_n(t)}\nonumber\\
&~~~~~~~~~~~~~~~~~~+
\average{R_m(t)R_n(t)(R_p\circ R_q)}-\average{R_m(t) R_n(t)}\average{R_p\circ R_q}
\big)\nonumber\\
& = \frac{1}{2}\sum_{p,q} C_{p,q}\Big(
2\average{R_p R_m(t)}\average{R_q R_n(t)}+
2\average{R_p R_n(t)}\average{R_q R_m(t)}\nonumber\\
&~~~~~~~~~~~~~~~~~~+
2\average{R_m(t) R_p}\average{R_n(t) R_q}+
2\average{R_m(t) R_q}\average{R_n(t) R_p}
\Big)\nonumber\\
& = \frac{1}{2}\sum_{p,q} C_{p,q}\Big(
4\average{\left\{R_p~,~R_m(t)\right\}}\average{\left\{R_q~,~R_n(t)\right\}}\nonumber\\
&~~~~~~~~~~~~~~~~~~+
 \average{\left[R_p~,~R_m(t)\right]}\average{\left[R_q~,~R_n(t)\right]}
\nonumber\\
&~~~~~~~~~~~~~~~~~~+
4\average{\left\{R_q~,~R_m(t)\right\}}\average{\left\{R_p~,~R_n(t)\right\}}\nonumber\\
&~~~~~~~~~~~~~~~~~~+
 \average{\left[R_q~,~R_m(t)\right]}\average{\left[R_p~,~R_n(t)\right]}
\Big)\nonumber\\
& = \frac{1}{2}\sum_{p,q} \Big(
4\sigma^{t0}_{m,p} C_{p,q} (\sigma^{t0})^{T}_{q,n} -
\Omega^{t0}_{m,p} C_{p,q} (\Omega^{t0})^{T}_{q,n} \nonumber\\&
~~~~~~~~~~~+
4\sigma^{t0}_{m,q} C_{q,p} (\sigma^{t0})^{T}_{p,n} -
\Omega^{t0}_{m,q} C_{q,p} (\Omega^{t0})^{T}_{p,n}
\Big)\nonumber\\
& = 
\left(4\sigma^{t0} C (\sigma^{t0})^T -
\Omega^{t0} C (\Omega^{t0})^T\right)_{m,n}.
\end{align}
Here, from the first to the second equality we use the fact that the average of the SLD is zero, from the third equality to the fourth we use the Wick's theorem, and from the fifth to the sixth one we use the fact that $C$ is a symmetric matrix, and $\sigma^{t0}_{i,j}=(\sigma^{t0})^T_{j,i}$, and $\Omega^{t0}_{i,j}=-(\Omega^{t0})^T_{j,i}$.
Notice that the matrix inside parenthesis in the last line is symmetric, and therefore ${\rm Corr}(\Lambda_0,\Sigma(t)_{m,n}) = {\rm Corr}(\Lambda_0,R_m(t)R_n(t))$.
Finally, since this identity is true for any element of $\Sigma(t)$, we have:
\begin{align}
{\rm Corr}(\Lambda_0,\Sigma(t)) = 4\sigma^{t0} C (\sigma^{t0})^T - \Omega^{t0} C (\Omega^{t0})^T,
\end{align}
which together with \eqref{eq-On-the-one-hand} completes our proof.
%
\section{From quadratic Lindbladian master equation to the master equation for the CM}\label{App:C}
Given a Lindbladian master equation that acts on the density matrix, how can we build up the equivalent master equation that operates on the moments (specifically on the first, and the second moments). 
To this aim, let us consider the most generic Lindbladian ME:
\begin{align}
	{\dot \rho} = -i[H~,~\rho] + \sum_{k=1}^{m}\left(L_k \rho L_k^{\dagger} - \frac{1}{2}\left\{L_k^{\dagger} L_k~,~\rho\right\}\right),
\end{align}
with the quadratic Hamiltonian $H = \frac{1}{2} R^TG R$,
and the linear Lindbladian operators $L_k = c_k^T R$.
In the Heisenberg picture, for any observable $O$---that does not depend explicitly on time---this reads as
\begin{align}
{\dot O} = i[H~,~O] + \sum_{k=1}^{m}\left(L_k^{\dagger} O L_k - \frac{1}{2}\left\{L_k^{\dagger} L_k~,~O\right\}\right).
\end{align}
\subsection{First moments}
We start by the first moments. Namely, for the $R_j$ element, we can calculate the terms that appear above, individually. For the Hamiltonian part we have
\begin{align}
	i[H~,~R_j] 
	= \frac{i}{2}G_{l,m} [R_l R_m~,~R_j] = \frac{i}{2}G_{l,m} (\Omega_{lj} R_m + \Omega_{mj} R_l) = -i(\Omega G R)_j.
\end{align}
The dissipation part of the dynamics has two contributions. The first part can be written as (we drop the index of the Lindbladian operators to avoid confusion, but we shall reconsider them later on):
\begin{align}
	L^{\dagger} R_j L
	& = R^T c^* R_j c^T R = c^*_l c_m R_l R_j R_m  = c^*_l c_m R_l R_m R_j ~ + ~ c^*_l c_m \Omega_{jm} R_l \nonumber\\
	& = c^*_l c_m R_l R_m R_j ~ + ~ (\Omega c c^{\dagger} R)_j.
\end{align}
In the same manner, we can deal with the second part:
\begin{align}
	-\frac{1}{2}\left\{L^{\dagger} L~,~R_j\right\}
	& = -\frac{1}{2}\left( R^T c^* c^T R R_j ~+~ R_j R^T c^* c^T R   \right)\nonumber\\
	& = -\frac{1}{2}\left( c_l^* c_m R_l R_m R_j ~+~ c_l^* c_m R_j R_l R_m  \right)\nonumber\\
	& = -\frac{1}{2}\left( c_l^* c_m R_l R_m R_j ~+~ c_l^* c_m R_l R_j R_m ~+~ c_l^* c_m \Omega_{jl} R_m \right)\nonumber\\
	& = -\frac{1}{2}\left( c_l^* c_m R_l R_m R_j ~+~ c_l^* c_m R_l R_m R_j ~+~ c_l^* c_m \Omega_{jm} R_l ~+~ c_l^* c_m \Omega_{jl} R_m \right)\nonumber\\
	& = - c_l^* c_m R_l R_m R_j - \frac{1}{2} (\Omega c^* c^T R ~+~ \Omega c c^{\dagger} R)_j.
\end{align}
Putting the two parts together, we have:
\begin{align}
	L^{\dagger} R_j L - \frac{1}{2}\left\{L L^{\dagger}~,~R_j\right\}
	& = \frac{1}{2} (\Omega c c^{\dagger} R ~-~ \Omega c^* c^T R)_j = \left[i\Omega~{\rm Im}(c c^{\dagger}) R\right]_j.
\end{align}
By considering all of the Lindbladian operators, in a more compact form we have:
\begin{align}
	\dot{R} 
	& = \left(-i\Omega G~ + i\Omega \sum_{k=1}^{m}~{\rm Im}(c_k c_k^{\dagger})\right)R = AR,
\end{align}
with $A \coloneqq -i\Omega\left(G~ - {\rm Im}(C C^{\dagger}) \right)$, and $C \coloneqq (c_1^T; c_2^T; \dots c_m^T)^T $ is a $2N\times m$ matrix containing the dissipation coefficients.
\subsection{Second moments}
We choose the same procedure to deal with the second moments. To begin with, we explore the Hamiltonian term:
\begin{align}
	i[H~,~R_j R_k] 
	& = i([H~,~R_j] R_k + R_j[H~,~R_k]) = -i(\Omega G R)_j R_k -i R_j (\Omega G R)_k\nonumber\\
	& = -i(\Omega G R R^T)_{jk} -i (R R^T G^T \Omega^T)_{jk}.
\end{align}
Moreover, for the dissipators we have:
\begin{align}
	L^{\dagger} R_j R_k L
	& = c^*_m c_n R_m R_j R_k R_n = c^*_m c_n R_m R_n R_j R_k + c^*_m c_n R_m R_k \Omega_{jn} + c^*_m c_n R_m R_j \Omega_{kn}.
\end{align} 
Let us deal with the two terms appearing in the anticommutator separately. The first one reads:
\begin{align}
	-\frac{1}{2} L^{\dagger} L R_j R_k
	& = -\frac{1}{2} c_m^* c_n R_m R_n R_j R_k,
\end{align}
and finally the second one reads as:
\begin{align}
-\frac{1}{2} R_j R_k L^{\dagger} L
& = -\frac{1}{2}~c_m^* c_n R_j R_k R_m R_n \nonumber\\
& = -\frac{1}{2}\left\{c_m^* c_n R_j R_m R_n R_k + c_m^* c_n R_j R_m \Omega_{kn} + c_m^* c_n R_j R_n \Omega_{km}\right\}\nonumber\\
& = -\frac{1}{2}\big\{c_m^* c_n R_m R_n R_j R_k + c_m^* c_n R_m R_k \Omega_{jn} + c_m^* c_n R_n R_k \Omega_{jm} + c_m^* c_n R_j R_m \Omega_{kn} \nonumber\\
&~~~~~~~~~+ c_m^* c_n R_j R_n \Omega_{km}\big\}.
\end{align}
Putting the last three expressions together in the dissipation part of the master equation yields:
\begin{align}
	L^{\dagger} R_j R_k L - \frac{1}{2}\left\{ L^{\dagger} L~ ,~R_j R_k \right\}
	& = \Omega_{jn} c^*_m c_n R_m R_k + \Omega_{kn} c_n c^*_m R_m R_j
	  -\frac{1}{2}\big\{\Omega_{jm} c_m^* c_n R_n R_k\nonumber\\
	& + \Omega_{jn} c_n c_m^* R_m R_k + \Omega_{km} c_m^* c_n R_j R_n + \Omega_{kn} c_n c_m^* R_j R_m
	  \big\}\nonumber\\
	& = \left[\Omega c c^{\dagger} R R^T\right]_{jk} - \left[R R^T c^* c^{T} \Omega \right]_{jk} + \left[\Omega c^* c^{T} \Omega \right]_{jk}\nonumber\\
	&   -\frac{1}{2}\left[\Omega c^* c^{T} R R^T\right]_{jk} -\frac{1}{2}\left[\Omega c c^{\dagger} R R^T\right]_{jk} \nonumber\\
	& + \frac{1}{2}\left[R R^T c c^{\dagger} \Omega \right]_{jk} + \frac{1}{2}\left[R R^T c^* c^{T} \Omega \right]_{jk}\nonumber\\
	& = \frac{1}{2}\left[\Omega c c^{\dagger} R R^T - \Omega c^* c^{T} R R^T \right]_{jk} + \frac{1}{2}\left[R R^T c c^{\dagger} \Omega - R R^T c^* c^{T} \Omega \right]_{jk}~\nonumber\\
	& + \left[\Omega c^* c^{T} \Omega \right]_{jk}\nonumber\\
	& = i\left[\Omega {\rm Im} (c c^{\dagger}) R R^T \right]_{jk} + i\left[R R^T {\rm Im} (c c^{\dagger}) \Omega \right]_{jk}
	+ \left[\Omega c^* c^{T} \Omega \right]_{jk}.
\end{align}
If we add this to the same dissipation part, but with the elements $j \leftrightarrow k$, and divide by two, we have:
\begin{align}
	L^{\dagger} \Sigma_{jk} L - \frac{1}{2}\left\{ L^{\dagger} L~,~\Sigma_{jk} \right\} 
	& = i\left[\Omega~{\rm Im} (c c^{\dagger}) \Sigma \right]_{jk} + i\left[\Sigma~{\rm Im} (c c^{\dagger}) \Omega \right]_{jk}
+ \left[\Omega~{\rm Re}(c c^{\dagger}) \Omega \right]_{jk}.
\end{align}
Thus, for the second moments, after adding up all dissipators, we will have the following master equation:
\begin{align}
	\frac{d \Sigma}{dt} = A \Sigma + \Sigma A^T + D,
\end{align}
with $A = -i\Omega(G - {\rm Im}(C C^{\dagger}))$, and $D = \Omega~{\rm Re}(C C^{\dagger}) \Omega$, with the same definition of $C$ that we have for first moments, i.e., $C \coloneqq (c_1^T; c_2^T; \dots c_m^T)^T $.
\end{document}